\documentclass[twocolumn]{aastex631}

\usepackage{ifthen}
\usepackage{afterpage}

\newcommand{\comment}[1]{}

\newcommand{\change}[1]{{\color{black} #1}}
\newcommand{\changeenv}[1]{\color{black}}

\newcommand{\changetwo}[1]{{\color{black} #1}}

\usepackage{amsmath}
\usepackage{natbib}
\usepackage{appendix}
\usepackage{gensymb}
\usepackage{wasysym}
\usepackage{bm}
\usepackage{tablefootnote}

\providecommand{\adsurl}[1]{\href{#1}{ADS}}

\DeclareSymbolFont{UPM}{U}{eur}{m}{n}
\DeclareMathSymbol{\umu}{0}{UPM}{"16}
\let\oldumu=\umu
\renewcommand\umu{\ifmmode\oldumu\else$\oldumu$\fi}
\newcommand\micro{\umu}

\newcommand\microns{\micro m}

\newcommand\lmax{$l_{\rm max}$}

\comment{\hfill\textbf{DRAFT of {\today} \now}.}
\shorttitle{WASP-43b Eclipse Mapping}
\shortauthors{Challener et al.}

\begin{document}

\title{Latitudinal Asymmetry in the Dayside Atmosphere of WASP-43b}
\author[0000-0002-8211-6538]{Ryan C. Challener}
\affiliation{Department of Astronomy, Cornell University, 122 Sciences Drive, Ithaca, NY 14853, USA}

\author[00000-0003-4408-0463]{Zafar Rustamkulov}
\affiliation{Department of Earth and Planetary Sciences, Johns Hopkins University, 3400 N. Charles Street, Baltimore, MD 21218, USA}

\author[0000-0002-3052-7116]{Elspeth K.H. Lee}
\affiliation{Center for Space and Habitability, University of Bern, Gesellschaftsstrasse 6, CH-3012 Bern, Switzerland}

\author[0000-0002-8507-1304]{Nikole Lewis}
\affiliation{Department of Astronomy, Cornell University, 122 Sciences Drive, Ithaca, NY 14853, USA}

\author[0000-0001-6050-7645]{David K. Sing}
\affiliation{Department of Physics and Astronomy, Johns Hopkins University, 3400 N. Charles Street, Baltimore, MD 21218, USA}
\affiliation{Department of Earth and Planetary Sciences, Johns Hopkins University, 3400 N. Charles Street, Baltimore, MD 21218, USA}

\author[0000-0001-7058-1726]{Stephan M. Birkmann}
\affiliation{European Space Agency, European Space Astronomy Centre, Camino Bajo del Castillo s/n, E-28692 Villanueva de la Cañada, Madrid, Spain}

\author[0000-0001-7866-8738]{Nicolas Crouzet}
\affiliation{Leiden Observatory, Leiden University, P.O. Box 9513, 2300 RA Leiden, The Netherlands}

\author[0000-0001-9513-1449]{N\'estor Espinoza}
\affiliation{Space Telescope Science Institute, 3700 San Martin Drive, Baltimore, MD 21218, USA}
\affiliation{Department of Physics and Astronomy, Johns Hopkins University, 3400 N. Charles Street, Baltimore, MD 21218, USA}

\author[0000-0003-0192-6887]{Elena Manjavacas}
\affil{AURA for the European Space Agency (ESA), ESA Office, Space Telescope Science Institute, 3700 San Martin Drive, Baltimore, MD 21218, USA}
\affiliation{Department of Physics and Astronomy, Johns Hopkins University, 3400 N. Charles Street, Baltimore, MD 21218, USA}

\author[0000-0001-5254-6740]{Natalia Oliveros-Gomez}
\affiliation{Department of Physics and Astronomy, Johns Hopkins University, 3400 N. Charles Street, Baltimore, MD 21218, USA}

\author[0000-0003-3305-6281]{Jeff A. Valenti}
\affiliation{Space Telescope Science Institute, 3700 San Martin Drive, Baltimore, MD 21218, USA}

\author[0009-0006-2395-6197]{Jingxuan Yang}
\affiliation{Atmospheric, Oceanic and Planetary Physics, Department of Physics, University of Oxford, Oxford OX1 3PU, UK}

% Bug in AASTeX v. 6.3 requires this
%\nocollaboration{0}

% Suggested: 250 words or less.
\begin{abstract}

We present two-dimensional near-infrared temperature maps of the canonical hot Jupiter WASP-43b using a phase-curve observation with JWST NIRSpec/G395H.
From the white-light planetary transit, we improve constraints on the planet's orbital parameters and measure a planet-to-star radius ratio of \change{$0.15883^{+0.00056}_{-0.00053}$}.
Using the white-light phase curve, we measure a longitude of maximum brightness of $6.9^{+0^\circ.5}_{-0^\circ.5}$ east of the substellar point and a phase-curve offset of \change{$10.0^{+0^\circ.8}_{-0^\circ.8}$}.
We also find an $\approx4\sigma$ detection of a latitudinal hotspot offset of \change{$-13.4^{+3^\circ.2}_{-1^\circ.7}$}, the first significant detection of a non-equatorial hotspot in an exoplanet atmosphere. 
We show that this detection is robust to variations within planetary parameter uncertainties, but only if the transit is used to improve constraints\change{, showing the importance of transit observations to eclipse mapping}.
Maps retrieved from the NRS1 and NRS2 detectors are similar, with hotspot locations consistent between the two detectors at the $1\sigma$ level.
Our JWST data show brighter (hotter) nightsides and a dimmer (colder) dayside at the shorter wavelengths relative to fits to \textit{Spitzer} 3.6 and 4.5 \microns\ phase curves. 
Through comparison between our phase curves and a set of general circulation models, we find evidence for clouds on the nightside and atmospheric drag or high metallicity reducing the eastward hotspot offset.

\end{abstract}


\section{Introduction}

Eclipses of transiting exoplanets, where the planet is blocked by the star, enable multidimensional characterization of exoplanet atmospheres \citep{RauscherEtal2007apjEclipseMapping, CowanFujii2018haexMappingReview}.
As the stellar disk covers and uncovers the planet, the planet's spatial brightness distribution is imprinted on the shape of the eclipse light-curve ingress and egress.
Eclipses constrain smaller-scale and latitudinal brightness variations, unlike phase-curve observations, which are limited to large-scale longitudinal structure and cannot constrain latitudinal variation.
For photometric or white-light eclipses, this enables two-dimensional brightness (temperature) mapping of atmospheres.
Spectroscopic light curves, in addition to two-dimensional horizontal information, constrain vertical thermal structure and composition.
Thus, spectroscopic eclipse observations of sufficient precision constrain the three-dimensional temperatures and composition of exoplanet atmospheres \citep[e.g.,][]{MansfieldEtal2020mnrasEigenspectraMapping, ChallenerRauscher2022ajThERESA}.

Currently, three planets have been eclipse-mapped.
First, HD~189733~b was mapped by stacking seven \textit{Spitzer} InfraRed Array Camera 8 \microns\ photometric eclipses and a partial phase curve \citep{MajeauEtal2012apjlHD189Map, DeWitEtal2012aaHD189Map, RauscherEtal2018ajMap, ChallenerRauscher2022ajThERESA}.
These data were able to constrain the longitudinal hotspot offset in the planet's atmosphere, but the relatively low precision and lack of spectroscopic information limited mapping capabilities.
Using a single JWST NIRSS/SOSS eclipse from the Transiting Exoplanet Community Early Release Science Program (ID: 1366), \cite{CoulombeEtal2023natWASP18b} created a white-light map of WASP-18b with much higher precision, showing the data could be explained by either a traditional hotspot or a dayside temperature ``plateau'', where both models were consistent with significant atmospheric drag.
In the same program, \cite{HammondEtal2023apjW43bMIRImap} used a MIRI/LRS phase-curve observation of WASP-43b to create a white-light eclipse map, showing a small longitudinal hotspot offset and the first observational evidence for latitudinal temperature gradients in an exoplanet atmosphere. 

Discovered in 2011 \citep{HellierEtal2011aapWASP43b}, WASP-43b is a hot Jupiter orbiting its K7V host star every 0.81 days.
The planet-to-star brightness ratio of WASP-43b is higher than most other systems, making it an excellent target for atmospheric characterization through emission observations.
At 2.052 $M_{\textrm{J}}$, WASP-43b has one of the few high-gravity exoplanet atmospheres that can be characterized.
It bridges the gap between planets and Brown Dwarfs, probing how scale height affects atmospheric processes. 
\cite{BlecicEtal2014apjWASP43b} used \textit{Spitzer} InfraRed Array Camera eclipse observations to rule out a strong atmospheric thermal inversion.
\cite{StevensonEtal2014sciWASP43bphasecurve}, using \textit{Hubble} Wide Field Camera 3 ($1.1 - 1.7$ \microns) phase-curve observations, presented the phase-resolved thermal structure of the planet, which showed a non-inverted temperature profile at all phases, a hotspot offset of $12.3 \pm 1^\circ.0$ east of the substellar point, and an asymmetric phase curve.
\cite{StevensonEtal2017ajWASP43bPhaseCurve} analyzed \textit{Spitzer} phase curves at both $3.6$ and $4.5$ \microns, finding \change{longitudinal} hotspot offsets of $12.2 \pm 1^\circ.7$ and $21.1 \pm 1^\circ.8$, respectively.
Further analyses of this \textit{Spziter} 4.5 \microns\ phase curve with updated methods found a smaller and more uncertain \change{longitudinal} hotspot offset \citep{MendoncaEtal2018ajWASP43b, MorelloEtal2019ajW43bPhaseCurves, MayStevenson2020ajNewBLISS, BellEtal2021mnrasSpitzer45um}, and a joint analysis with two additional 4.5 \microns\ phase curves found the \change{longitudinal} hotspot offset to be $10.4 \pm 1^\circ.8$, consistent with 3.6 \microns\ \citep{MurphyEtal2023ajWASP43bNoVariability}.
Others demonstrated the multidimensional information content of phase curves with application to WASP-43b \citep{FengEtal2020aj2DRetrievals, IrwinEtal2020mnrasTwoPlusD, ChubbMin2022aapARCiS3D,
yang2023testing}.

Here, we present an eclipse-mapping analysis of WASP-43b using a JWST \citep{GardnerEtal2006ssrJWST} NIRSpec/G395H \citep{JakobsenEtal2022aapNIRSpec} phase-curve observation covering $2.87 - 5.18$ \microns.
We present a high-precision full-bandpass 2D brightness-temperature map of the entire planet, as well as brightness temperature maps over the NRS1 (2.87 -- 3.72 \microns) and NRS2 (3.82 -- 5.18 \microns) detectors.
In Section \ref{sec:data} we describe the observation and data reduction, in Section \ref{sec:transit} we use the white-light transit to refine planetary parameters, in Section \ref{sec:methods} we outline our mapping methods, in Section \ref{sec:whitelight} we present our mapping analysis of the white-light phase curve, in Section \ref{sec:2channel} we show maps fit to the NRS1 and NRS2 light curves, in Section \ref{sec:gcms} we compare our data and maps against theoretical predictions, and in Section \ref{sec:conclusions} we lay out our conclusions.

\section{Observation and Data Reduction}
\label{sec:data}

We use the JWST NIRSpec/G395H phase-curve observation of WASP-43b in program 1224 (PI: Stephan Birkmann).
The observation began on May 14, 2023 at 9:03:30 UT and ran for 28.57 hours, beginning shortly before an eclipse of the planet, capturing the planet's full rotational light curve, including two eclipses and one transit.

We use FIREFLy \citep{RustamkulovEtal2022apjlFIREFLy, RustamkulovEtal2023natWASP39b} to extract the spectrophotometry from the raw $\texttt{uncal.fits}$ files. The pipeline performs calibrations and up-the-ramp fitting using the $\texttt{jwst}$ pipeline, treats bad pixels and cosmic rays, subtracts 1/$f$ noise at the group-level, corrects for spatial shifts of the trace, and extracts the spectrophotometry. The cleaning thresholds for anomalous pixels and cosmic ray hits are tuned to ensure minimal impact on the extracted stellar fluxes. We extract the spectrophotometry from a best-fit spectral trace, and qualitatively optimize the extraction width to minimize the point-to-point scatter. We fit and remove a small linear trend in each wavelength bin using the flux level inside both eclipses. More complex systematics models are not warranted by the high-quality data, which is also the case for a similar phase curve observation of WASP-121 b in the same mode \citep{MikalEvansEtal2023apjlW121}. In particular, neither the white-light nor the spectroscopic light curves favor systematics models that include higher-order polynomials or telescope jitter at the 1$\sigma$-level.

We assume photon-limited uncertainties, and calculate a ``white-light'' light curve as an inverse-variance-weighted sum of the spectroscopic light curves. 
\change{We do not include an offset between the NRS1 and NRS2 detectors \citep[e.g.,][]{MayEtal2023apjlGJ1132}, and preliminary analysis of the emission spectra, which will be presented in follow-up work, shows no significant evidence for an offset.}
All the mapping analyses in this work operate on these detrended data.

\change{
The data used in this paper can be found on the Mikulski Archive for Space Telescopes: \dataset[10.17909/n3ac-1s50]{http://dx.doi.org/10.17909/n3ac-1s50}.
}

\section{System Parameters}
\label{sec:transit}

Eclipse mapping is highly sensitive to system parameters, particularly the orbital inclination $i$ \citep{ChallenerRauscher2023ajNullSpace}.
\changetwo{Here, we use our data to refine the system parameters. 
Because our mapping method (Section \ref{sec:methods}) cannot co-fit the system parameters and map,} we divide the data into two sections: (1) the transit (60078.910 - 60078.975 BMJD), which we fit to refine the system parameters before performing the mapping fit, and (2) the phase curve and eclipses, which we use for the mapping fit (see Section \ref{sec:whitelight}).
\changetwo{In this way, we conservatively avoid fitting to the same data twice (i.e., in both orbital parameter estimation and mapping) and we avoid potential biases introduced by fitting a non-mapping model to the eclipses \citep{BellEtal2023natasWASP43bMIRI, HammondEtal2023apjW43bMIRImap}.
We discuss the system parameters and their effects on mapping in more detail in Sections \ref{sec:methods} and \ref{sec:verification}.}

\begin{figure}
    \includegraphics[width=3.25in]{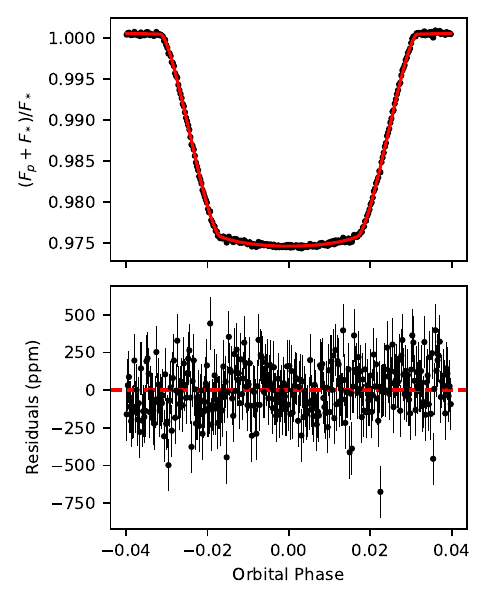}
    \caption{White-light transit and fit used to determine the system parameters. The residuals show some evidence of correlated noise, particularly during ingress, which could indicate asymmetries in the planetary limbs, although WASP-43 is a K star with some evidence of long-term ($>15$ days) rotational modulation in the light curve \citep{HellierEtal2011aapWASP43b}, so starspots could also contribute to correlated noise. Detailed modeling of the transit will be explored in follow-up work.\label{fig:transit}}
\end{figure}

We use BATMAN \citep{Kreidberg2015paspBATMAN} to model the transit, with a quadratic limb-darkening model.
We treat planetary nightside emission contamination as a constant plus a sinusoid with a minimum at mid-transit (sinusoids are commonly used to fit phase-curve variation, e.g., \citealp{StevensonEtal2017ajWASP43bPhaseCurve, BellEtal2023natasWASP43bMIRI}).
In total we fit to eight parameters: mid-transit time $t_0$, planet-to-star radius ratio $R_p/R_*$, orbital semi-major axis $a/R_*$, orbital inclination $i$, quadratic limb-darkening parameters \change{$q_1$} and \change{$q_2$ \citep{Kipping2013mnrasLimbDarkLaws}}, planetary nightside flux constant $F_c$, and planetary nightside flux sinusoidal amplitude $F_a$.
\change{Note that we tested a model which fit separately for $t_0$ and the time of minimum nightside emission, but that model only marginally improved the Bayesian Information Criterion (BIC, \citealp{Raftery1995BIC}) by 1.3 (1.9:1 relative model preference) and resulted in a non-physical negative \changetwo{constant} nightside emission \changetwo{term} and phase variation decreasing away from \changetwo{near} mid-transit.}
We use two-sided Gaussian priors on $t_0$, $i$, and $a/R_*$ from the eclipse-mapping fit of the WASP-43b MIRI/LRS phase-curve observation in \cite{HammondEtal2023apjW43bMIRImap}.
Using MC3 \citep{CubillosEtal2017ajRedNoise}, we determine the best-fitting model and explore the parameter space with Markov-chain Monte Carlo (MCMC) to estimate parameter uncertainties.
The data, best-fitting model, and model residuals are shown in Figure \ref{fig:transit}.
Our fitted and assumed parameters are listed in Table \ref{tab:params}.
\change{In general, we reduce uncertainties on fitted parameters by approximately a factor of 2 over \cite{HammondEtal2023apjW43bMIRImap}.}

We note that our measured transit time $t_0$ is approximately 3$\sigma$ \change{(2.8 seconds)} earlier than the $t_0$ prior we adopt from \cite{HammondEtal2023apjW43bMIRImap}. 
To ensure the prior is not enforcing a poor fit to the data, we refit the transit without a prior on $t_0$ and find that the time of transit shifts by 1.6$\sigma$, or 1.4 seconds. 
Thus, we are confident that the difference between our prior and measurement is not problematic.
Regardless, we later show in Section \ref{sec:verification} that our results are not impacted by $t_0$ variations of this magnitude.

\begin{table*}[]
\changeenv{}
    \centering
    \begin{tabular}{l|r|r}
         Parameter & Value & Prior\\
         \hline
         \hline
         Assumed Parameters\\
         \hline
         Orbital period $P$ (days) & 0.8134741 \\
         Eccentricity $e$ & 0 \\
         Stellar mass $M_*$ ($M_\odot$) & 0.6916 \\
         Stellar radius $R_*$ ($R_\odot$) & 0.665 \\
         Stellar effective temperature $T_*$ (K)& 4520 \\
         Planet mass $M_p$ ($M_{\textrm J}$) & 2.052  \\
         \hline
         Fitted Parameters\\
         \hline
         Planet radius $R_p$ ($R_*$) & $0.15883^{+0.00056}_{-0.00053}$ & $\mathcal{U}(0.15, 0.17)$\\
         Inclination $i$ ($^\circ$) & $82.155^{+0.027}_{-0.027}$ & $\mathcal{N}(82.105, -0.051, 0.050)$\\
         Semimajor axis $a$ ($R_*$) & $4.8767^{+0.0068}_{-0.0069}$ & $\mathcal{N}(4.859, -0.012, 0.012)$ \\
         Transit time $t_0$ (BMJD) & $55934.2922503^{+0.0000063}_{-0.0000062}$ & $\mathcal{N}(55934.292283, -0.000011, 0.000011)$\\
         Limb darkening $q_1$ & $0.044^{+0.011}_{-0.011}$ & $\mathcal{U}(0.0, 1.0)$\\
         Limb darkening $q_2$ & $0.27^{+0.24}_{-0.17}$ & $\mathcal{U}(0.0, 1.0)$ \\
         Planet emission constant $F_c$ & $0.0083^{+0.0060}_{-0.0059}$ & $\mathcal{U}(-0.1, 0.1)$\\
         Planet sinusoidal amplitude $F_a$ & $-0.0080^{+0.0061}_{-0.0061}$ & $\mathcal{U}(-1.0, 1.0)$ \\
         \hline
    \end{tabular}
    \caption{WASP-43 system parameters from fitting to the white-light transit. A prior of $\mathcal{U}(a,b)$ denotes a uniform (non-informative) prior between $a$ and $b$, and a prior of $\mathcal{N}(\mu, a, b)$ denotes an asymmetric Gaussian prior with a mean of $\mu$, a lower 1$\sigma$ bound of $a$, and an upper 1$\sigma$ bound of $b$. Except $M_*$, the assumed parameters match those in \cite{HammondEtal2023apjW43bMIRImap}, and the Gaussian priors come from their mapping fit. $M_*$ is calculated from $a$ and $P$.}
    \label{tab:params}
\end{table*}

\section{Mapping Methods}
\label{sec:methods}

We used ThERESA \citep{ChallenerRauscher2022ajThERESA} for the mapping analysis.
ThERESA's capabilities have been demonstrated with comparisons to synthetic data \citep{ChallenerRauscher2022ajThERESA} and past analyses of HD 189733 b \citep{ChallenerRauscher2022ajThERESA}, a JWST NIRISS/SOSS eclipse of WASP-18b \citep{CoulombeEtal2023natWASP18b}, and a JWST MIRI/LRS phase curve of WASP-43b \citep{HammondEtal2023apjW43bMIRImap}.

ThERESA uses the ``eigenmapping'' approach \citep{RauscherEtal2018ajMap}.
We calculate the light curves generated by spherical harmonics up to degree \lmax, then apply principal component analysis to orthogonalize these light curves into a new basis set of ``eigencurves''.
The calculation of these eigencurves requires assuming orbital parameters; we use the orbital solution from fitting the transit (Section \ref{sec:transit}, Table \ref{tab:params}) with priors from a mapping analysis of a WASP-43b JWST MIRI/LRS observation of a full phase curve \citep{HammondEtal2023apjW43bMIRImap}.
\change{Note that these orbital parameters are statistically different than those in \cite{BellEtal2023natasWASP43bMIRI}, where an eclipse map was not fitted.}
We then rank these eigencurves by their variance (\change{i.e., their observability, or the total strength of the light-curve variation they generate; see \citealp{RauscherEtal2018ajMap}}) and fit the light curves as a weighted sum of the $N_E$ highest-variance eigencurves, along with a uniform mapping component $Y_0^0$ and a stellar correction parameter $s_{corr}$.
The full model is

\begin{equation}
    F_p / F_s = c_0 Y_0^0(t) + \sum_{i=1}^{N_E} c_i E_i(t) + s_{corr},
\end{equation}

\noindent
where the $c_i$ and $s_{corr}$ are free model parameters.
Each eigencurve has an associated ``eigenmap'', and the fitted light curves correspond to sums of the eigenmaps, using the same weights as the eigencurves.
We enforce a constraint that the fitted flux maps must be positive at all locations on the planet that are visible during the observation (for a full phase curve as analyzed here, the entire planet).
We include light-travel-time delay in our eigencurve calculation, and note that without this correction, the retrieved properties of the planet can change significantly.
We also tested a model including a linear slope and find a negligible ($<0.5\sigma$) change in the recovered hotspot \change{location} (see Section \ref{sec:whitelight}), so we are confident that removing a fitted slope (Section \ref{sec:data}) prior to fitting our mapping model does not impact our results.

We use MCMC through the MC3 package to explore the parameter space.
ThERESA tested all combinations of $l_{\rm max} \leq 6$ and $N_E \leq 15$, choosing the optimal combination using the \change{BIC}.
This chooses the appropriate amount of model complexity warranted by the quality of the data.

From the posterior distribution of eigenmap weights, we generate a posterior distribution of flux maps, from which we calculate a distribution of hotspot locations in latitude and longitude. 
Using a PHOENIX \citep{Husser2013} stellar spectrum (stellar temperature $T_* = 4500$ K, gravity log($g$) = 4.5, [Fe/H] = 0.0, and alpha element enhancement of [$\alpha$/H] = 0.0), and assuming the planet emits like a blackbody, we convert these flux maps, and the best-fitting flux map, into brightness temperatures. 
This distribution of brightness temperature maps gives an idea of the uncertainty on the measured map, but we note that the map uncertainty is limited by the patterns in the eigenmaps included in the fit.

\section{White-light Map}
\label{sec:whitelight}

\begin{figure}
    \centering
    \includegraphics[width=3.25in]{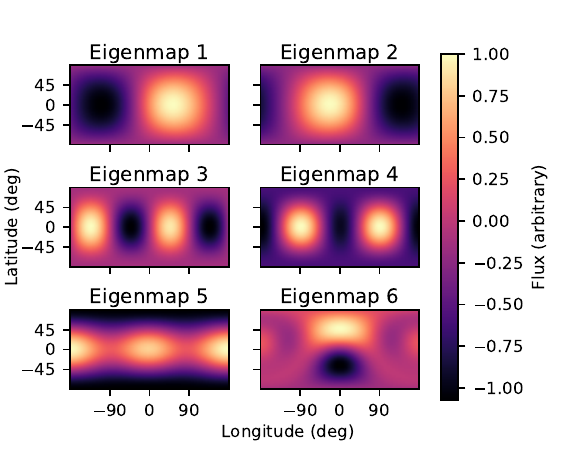}
    \caption{The six highest-variance (in light-curve space) eigenmaps used to model the white-light light curve. The maps are ordered from top left to bottom right in decreasing light-curve variance. The white-light data are modeled as a linear combination of these maps and, consequentially, the map may only contain these patterns. The map uncertainties are likewise restricted. The flux units are arbitrary, as the eigenmaps are scaled by free-parameter weights in the fitting process. For visual clarity, all eigenmaps have been normalized by their maxima.}
    \label{fig:emaps}
\end{figure}

The white-light light curve shows remarkable precision, requiring 3rd-degree harmonics and six eigencurves to be fit (Figure \ref{fig:emaps}).
The eigencurves and their associated eigenmaps are purely mathematical constructs of maximally-observable brightness patterns, so they are not linked to brightness patterns with a clear physical origin (e.g., expected dynamical patterns).
This has the benefit that the model is not reliant on assumptions, but can complicate interpretation of the model components.
However, we can examine their brightness patterns to understand how each eigenmap contributes to the overall fit and in which ways the model is flexible.

The data are dominated by out-of-eclipse samples, so the lower-number/higher-variance eigenmaps (1 -- 5) contain structures that have large signals in the phase curve.
Broadly, the first and second eigenmaps combine to set the day-night temperature contrast and the coarse longitudinal location of the planet's hotspot.
These components are detected at very high confidence ($>100\sigma$), as the full phase curve and two eclipses give significant information about these global-scale brightness patterns.
The third and fourth eigenmaps allow for finer adjustments to the longitudinal brightness distribution, roughly corresponding to higher-frequency sinusoids in the phase curve portions of the light curve.
The fifth eigenmap largely controls for the planet's equator-to-pole brightness gradient, and is latitudinally symmetric, but also contains longitudinal variation that shows up in the phase curve.
In contrast, the higher-number/lower-variance eigenmaps (6+) correspond to signals which are much stronger during eclipse than out-of-eclipse. 
The sixth eigenmap, which is required to adequately fit the data, introduces latitudinal asymmetry on the dayside of the planet, allowing the hotspot to deviate from the equator.
\change{All of the eigenmaps contain variation on a $\approx90^\circ$ or larger scale, so we can only retrieve large-scale features.}

\begin{figure*}
    \centering
    \includegraphics{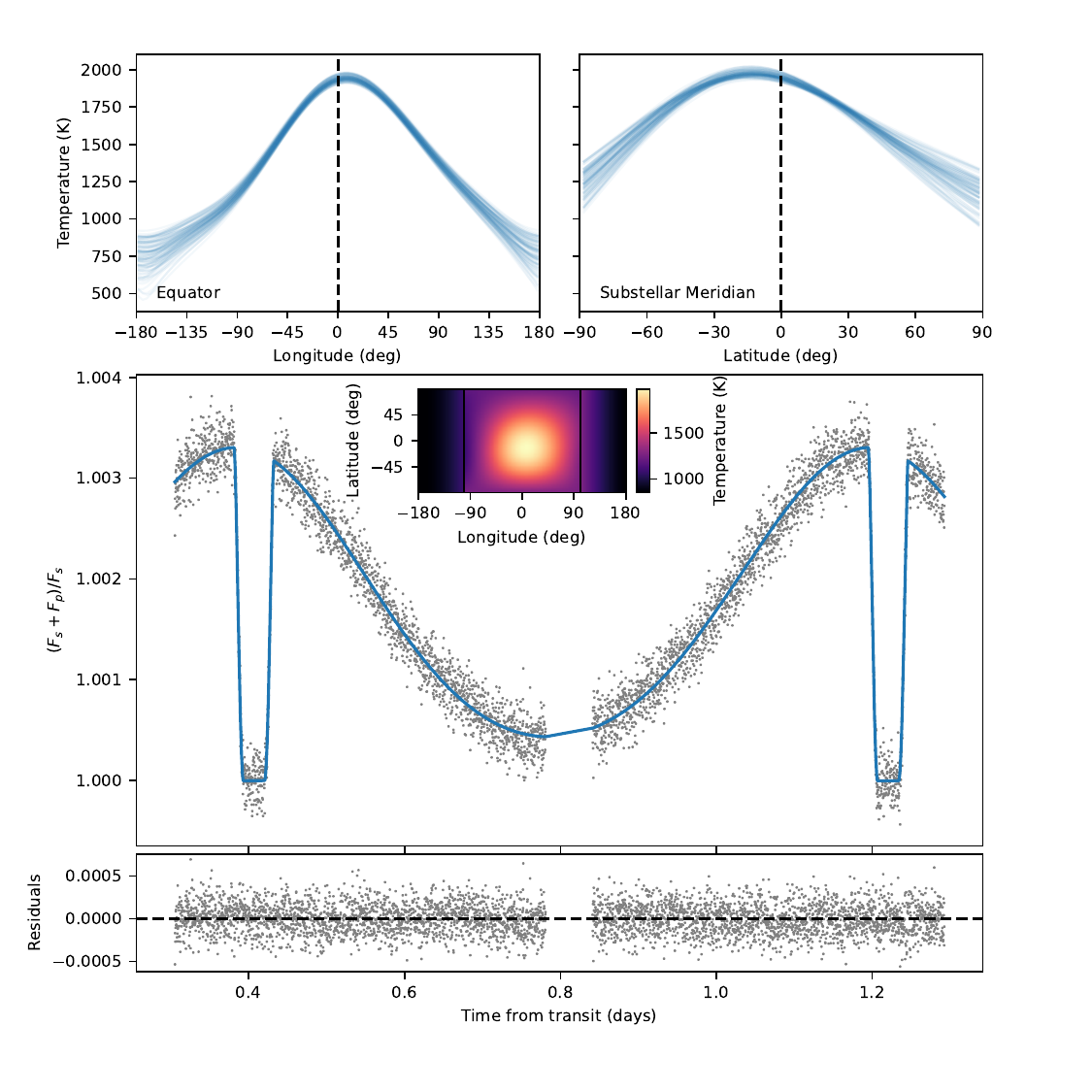}
    \caption{\changeenv{} {\bf Top:} A set of 100 of the posterior distribution temperature maps, sliced at the equator and substellar meridian, showing the range of plausible maps. The vertical black lines mark the substellar point. {\bf Middle:} The white-light light curve (gray), the best-fitting 2D eclipse-mapping model (blue), and the associated brightness temperature map (inset). The transit has been clipped from the dataset, as ThERESA cannot currently adequately model transit events, and the amount of emission mapping information lost is small. The best-fitting model uses 3rd-order spherical harmonics and 6 mapping components. The region between the vertical black lines on the inset is scanned by the stellar disk during either ingress, egress, or both. Outside these lines, the brightness temperatures are calculated from a cos(latitude)-weighted intensity average, to show that latitudinal information is unconstrained in that region. {\bf Bottom:} Residuals of the best-fitting model.}
    \label{fig:whitelight}
\end{figure*}

\begin{figure*}
    \centering
    \includegraphics[width=\textwidth]{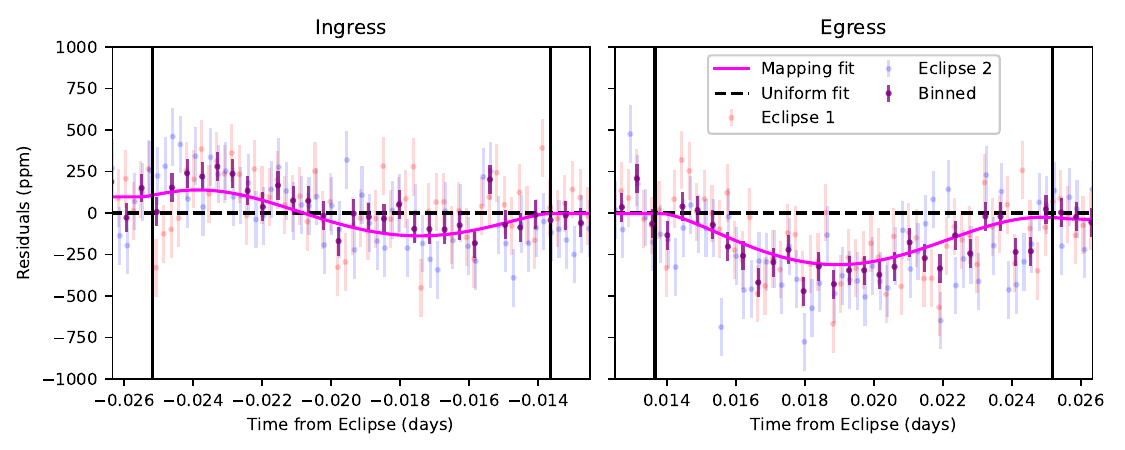}
    \caption{The difference between the white-light best-fitting map model and a uniform-planet model fit to the eclipses, to highlight the eclipse-mapping signal. \change{The purple data points show the data phase-folded and temporally binned by a factor of 4 for visual clarity.} The structure in the residuals shows signal in the data that cannot be fit by a uniform planet model (dashed black line) but is fit well by the mapping model (purple). Vertical black lines denote the beginning and end of ingress and egress. The uniform-planet model was fit to data within $\pm1$ hour of eclipse midpoint to reduce the impact of the phase-curve variation, which would be fit very poorly by a uniform model. The residuals, which are the difference between the uniform-planet model and the data, show clear structure that is fit well by the mapping model, and the shape of the residuals agree with the MIRI/LRS map \citep{HammondEtal2023apjW43bMIRImap}. The eclipses are consistent with each other.}
    \label{fig:ingressegress}
\end{figure*}

\change{The best-fitting map, the corresponding light-curve fit, the model residuals, and slices of the posterior distribution of temperature maps are shown in Figure \ref{fig:whitelight}}.
We achieve a reduced $\chi^2$ of 0.993. \change{(This suggests a slight 0.3\% overestimation of our uncertainties, but given the close proximity to 1.0, we elect to use the uncertainties as measured)}.
To highlight the eclipse-mapping signal, we also fit a uniform-brightness planet model to 2-hour chunks of data centered on the eclipses, to keep the fit from being overly influenced by the phase-curve variation, and compared this fit to the mapping model (Figure \ref{fig:ingressegress}).
The data are clearly inconsistent with the uniform-brightness planet, showing deviations in both ingress and egress that are fit well by the mapping model.
Additionally, the eclipses are consistent with each other, and the shape of the residuals (Figure \ref{fig:ingressegress}) closely match those seen with MIRI/LRS using a similar approach \citep{HammondEtal2023apjW43bMIRImap}.

The best-fitting map has a very large day-night temperature contrast, with a peak dayside temperature $\sim\,2000$ K dropping to an average of $\sim1000$ K on the nightside. 
The hottest point of the planet is $6.9^{+0^\circ.5}_{-0^\circ.5}$ east of the substellar point, indicating the presence of eastward winds.
This measurement is slightly lower but consistent with the \change{longitudinal} hotspot offset measured in \citep{HammondEtal2023apjW43bMIRImap}.
\change{We note that this uncertainty (and our other hotspot location uncertainties) is restricted by the patterns in the eiganmap basis set, and should be interpreted as the location of maximum brightness of large-scale features.
We cannot measure individual features at the resolution implied by this uncertainty.}

Here, we measure the location of maximum brightness as the ``hotspot offset'', which is a different measurement than the phase-curve offset reported in previous works, and consistency between those measurements is not necessarily expected \citep{ChallenerRauscher2023ajNullSpace}.
Therefore, for the purposes of comparison to past results, we compute the phase-curve offset of our best-fitting model and a posterior distribution of phase-curve offsets from our posterior distribution of map models by calculating the rotational angle that results in maximum brightness of the visible hemisphere.
For the white-light map, this offset is \change{$10.0^{+0^\circ.8}_{-0^\circ.8}$}. 
This is $\approx1\sigma$ west of the measurement at \textit{Spitzer} 3.6 \microns\ \citep[$12.2 \pm 1^\circ.7$,][]{StevensonEtal2017ajWASP43bPhaseCurve}.
The \textit{Spitzer} 4.5 \microns\  longitudinal offset varies with analysis method \citep{StevensonEtal2017ajWASP43bPhaseCurve, MendoncaEtal2018ajWASP43b, MorelloEtal2019ajW43bPhaseCurves, MayStevenson2020ajNewBLISS, BellEtal2021mnrasSpitzer45um}, but our measurement is consistent with a recent joint analysis of three phase curves \citep[$10.4 \pm 1^\circ.8$,][]{MurphyEtal2023ajWASP43bNoVariability}.

Our white-light map shows a latitudinal hotspot offset of \change{$-13.4^{+3^\circ.2}_{-1^\circ.7}$} (south) from the equator. \footnote{Note that in this work, we assume that the planet rotates on the same axis it orbits and that it eclipses north of the stellar equator, but it is equally likely that the planet eclipses south of the stellar equator. Changing this assumption would reflect our map across the planetary equator, but would not change the conclusion that the hotspot has a latitudinal offset. \change{\cite{HammondEtal2023apjW43bMIRImap} make the same assumption in their mapping of WASP-43b.}}
Our mapping model and light curve calculation include the effects of the tilt of the planet due to its orbital inclination; this latitudinal hotspot offset is not caused by the separation between the substellar point and the sub-observer point.
Thus far, eclipse-mapping hotspot locations are all equatorial or consistent with equatorial \citep{MajeauEtal2012apjlHD189Map, DeWitEtal2012aaHD189Map, RauscherEtal2018ajMap, ChallenerRauscher2022ajThERESA, CoulombeEtal2023natWASP18b, HammondEtal2023apjW43bMIRImap}. 
\cite{DeWitEtal2012aaHD189Map} found some evidence for a slight latitudinal hotspot offset in HD 189733b, but this detection was marginal and model-dependent, and further eigenmapping analyses of the same dataset found that latitudinal asymmetry was not necessary to adequately fit the data \citep{RauscherEtal2018ajMap, ChallenerRauscher2022ajThERESA}.
\change{\cite{HammondEtal2023apjW43bMIRImap} found a marginal latitudinal hotspot offset of $-10.7^{+4.^{\circ}1}_{-4.^{\circ}7}$ in their 2nd-order spherical-harmonic map of WASP-43b that agrees with our \change{latitudinal} offset in both magnitude and direction, although this was not presented as a detection due to concerns about systematic effects and model-dependent variation in this offset.}
Thus, our measurement represents the first significant detection of a latitudinal hotspot offset.

\subsection{Verification of the Latitudinal Hotspot Offset}
\label{sec:verification}

Formally, within the bounds of our optimal model (third degree spherical harmonic basis and six eigencurves/eigenmaps), the hotspot deviates from the equator by $4.2\sigma$.
Furthermore, the sixth eigenmap (ranked by eigencurve variance), which controls for hemisphere-scale asymmetric latitudinal flux variation on the dayside of the planet (Figure \ref{fig:emaps}), is detected at \change{5.8$\sigma$} confidence.
\change{However, as mentioned in Section \ref{sec:methods}, the uncertainties on our retrieved maps are limited by the six map components used in the fit, two of which control for latitudinal variation.
We ran additional models with up to 15 map components to understand the relationship between model complexity and the latitudinal hotspot offset.
While none of these additional components are necessary to fit the data well and their inclusion inflates our uncertainties, we find that the latitudinal hotspot offset is confidently detected ($>3\sigma$) for models with up to 12 components.
This highlights the robustness of our detection and the high quality of our data.}

As noted in Section \ref{sec:methods}, our mapping method assumes fixed values for the system parameters, to gain the benefits of orthogonal mapping components.
In reality, these system parameters have uncertainties that could impact the precision of our retrieved maps.
Previous analyses showed that system parameters have been known with sufficient certainty relative to the data precision to not significantly impact retrieved maps \citep{RauscherEtal2018ajMap, CoulombeEtal2023natWASP18b}.
However, given the high precision of our data, we test whether varying our assumed orbital parameter within their uncertainties changes the inferred latitudinal hotspot offset.

\begin{figure}
    \centering
    \includegraphics[width=3.25in]{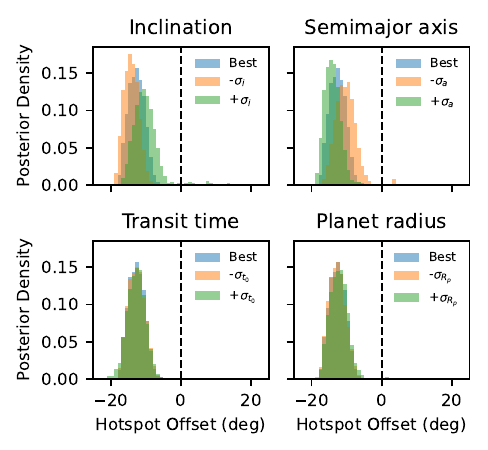}
    \caption{The effect of varying system parameters by $\pm1\sigma$ on the latitudinal hotspot offset. The vertical dashed black line marks the equator. The retrieved \change{latitudinal} hotspot offset remains consistent, and is non-equatorial.}
    \label{fig:orb}
\end{figure}

We follow the methods in \cite{CoulombeEtal2023natWASP18b}, and rerun the mapping analysis while adjusting each of $a$, $i$, $t_0$, and $R_p$ within 1$\sigma$ of their assumed values (Table \ref{tab:params}).
In all cases, we find the same optimal mapping model of 3rd-order spherical harmonics and six eigencurves.
That is, varying these parameters does not increase or decrease fit quality enough to warrant the removal or addition of new model parameters.
Varying the inclination and semimajor results in a small but noticeable change in the recovered latitudinal hotspot offset (Figure \ref{fig:orb}); adjusting the transit time and planet radius has no noticeable effect.
The measured latitudinal hotspot offsets are all consistent with each other and remain non-equatorial.
Thus, our measurement of a non-equatorial hotspot offset is robust to variations in system parameters.

We note that WASP-43b is one of the most well-studied hot Jupiters, and its parameters are well defined.
In this work, we leverage a previous JWST MIRI/LRS phase curve \citep{HammondEtal2023apjW43bMIRImap} and our JWST NIRSpec/G395H transit observation to achieve very precise orbital parameters.
We find that without both JWST observations, the inclination of the planet's orbit is not constrained enough to detect the planet's non-equatorial hotspot.
Precise determination of planetary parameters will be important to eclipse mapping in the JWST era. 
Studies that aim to map planets using only eclipses (not entire phase curves) should consider whether a transit observation is also necessary to sufficiently constrain orbital parameters in order to achieve their goals.

\subsection{Mechanisms for Creating Latitudinal Offsets}

In the presence of a pole-aligned magnetic field, an ionized atmosphere experiences a Lorentz drag that reduces east-west transport \citep{BeltzEtal2022ajActiveMagDrag3D}, and if the magnetic field is tilted relative to planetary rotation, atmospheric flows can be stably asymmetric, resulting in latitudinal hotspot offsets \citep{BatyginStanley2014apjObliqueMagFields}.
This interaction between the magnetic field and the atmosphere depends strongly on the magnetic field strength and the ionization state of the atmosphere, which is tied to temperature.
The brightness temperature of the dayside of WASP-43b peaks at $\sim2000$ K (Figure \ref{fig:whitelight}), lower than other similar short-period planets.
Thus, if the latitudinal hotspot offset is caused by magnetic interactions, the planet's magnetic field must be very strong.
\cite{YadavThorngren2017apjlMagFieldStrength} estimate hot Jupiter magnetic field strengths could reach as high as 50 -- 100 G, over an order of magnitude stronger than Jupiter's magnetic field. 

At the temperatures of WASP-43b's atmosphere, clouds may form, even on the dayside \citep{VenotEtal2020apjWASP43bGCMs, HellingEtal2020aapWASP43bCloudsHazes}.
Due to their opacity, clouds shift the height of the photosphere to lower pressures where the atmosphere is colder \citep{BlecicEtal2013apjWASP14b, StevensonEtal2014sciWASP43bphasecurve}.
Because we can only constrain large-scale spatial patterns, patchy clouds could significantly shift the apparent \change{latitudinal} location of the planet's hottest point, much like a similar effect predicted in longitudinal phase-curve offsets \citep{ParmentierEtal2021mnrasCloudyPhaseCurves}. 
However, CHEOPS, TESS, and HST observations showed no evidence of reflective clouds at short wavelengths \citep{FraineEtal2021ajW43bReflectionHST, ScandariatoEtal2022aapWASP43bNoCloudsDayside}, and a JWST MIRI/LRS spectrum of the dayside was consistent with a cloudless atmosphere \citep{BellEtal2023natasWASP43bMIRI}.
Furthermore, \textit{Spitzer} phase curves have shown no evidence of variability on WASP-43b on timescales of weeks to years \citep{MurphyEtal2023ajWASP43bNoVariability}, and the shape of our two eclipse are consistent with each other (Figure \ref{fig:ingressegress}). 
It is possible that the planet's dayside cloudiness is variable at a level that was undetectable prior to JWST and on a longer timescale than a single orbit, which could be explored in follow-up observations.
Detailed study of the emission and transmission spectra in our phase-curve observation, in combination with the same measurements in the MIRI/LRS observation \citep{BellEtal2023natasWASP43bMIRI}, will help elucidate the atmosphere's composition.

It has been shown that planets whose rotation axes do not align with their orbital axes (i.e., non-zero obliquity) can exhibit hotspots latitudinally offset from the equator and the sub-observer point \citep{Rauscher2017apjObliquity, RauscherEtal2023ajWarmJupiterMaps}.
However, assuming the system parameters listed in Table \ref{tab:params} and a tidal dissipation factor of $10^5$, appropriate for Jupiter \citep{GuillotEtal1996apjlCircularization}, WASP-43b has an extremely short tidal synchronization timescale of $\approx3000$ years.
Thus, we expect the planet's orbital and rotational periods to be synchronized, and for the planet's orbital and rotational axes to be aligned. 
If planetary obliquity is the source of the latitudinal hotspot offset in WASP-43b, the rotational axis must have become misaligned in the relatively recent past. 

Finally, high spatial resolution dynamical simulations have produced latitudinal offsets through intense storms \citep{ChoEtal2021apjlStorms}.
However, these simulations also produce quasi-periodic variability in their total planetary emission that has yet to be measured in hot Jupiter atmospheres \citep[e.g.,][]{KilpatrickEtal2020ajNoVariability, MurphyEtal2023ajWASP43bNoVariability}.
Both eclipses in our observation show similar morphology (Figure \ref{fig:ingressegress}), suggesting that the planet is not variable at a detectable level over the duration of one orbit, and MIRI/LRS observations at a different epoch agree.
If the latitudinal offset is due to variability, this variability is below current detection limits. 
Additional observations could place further constraints on the potential for variability.

\section{NRS1/NRS2 Mapping}
\label{sec:2channel}

We also fit mapping models to the data binned to the wavelengths of the NRS1 and NRS2 detectors, covering $2.65-3.72$ \microns\ and $3.82-5.17$ \microns, respectively.
This allows us to compare maps at different wavelengths while still maintaining a strong mapping signal.
We calculate these light curves and uncertainties using the same methods as the white-light light curve (Section \ref{sec:data}), although we find that this underestimates our uncertainties. 
Therefore, we increase our uncertainties to match the standard deviation of the data during total eclipse, where the planet signal is zero and, thus, the data are flat.
This amounts to uncertainties of 161 and 246 ppm for NRS1 and NRS2, respectively.
We find no evidence for significant temporal correlated noise in either detector (see Appendix \ref{sec:corrnoise}).

Figure \ref{fig:2channel} shows the light curves for each detector and the best-fitting models, with comparison to previous \textit{Spitzer} phase curves from \cite{StevensonEtal2017ajWASP43bPhaseCurve} at approximately the same wavelengths.
Note that the \textit{Spitzer} data has been binned temporally by a factor of 20 for visual clarity, reducing the true scatter of those data, while the JWST data are presented at observed cadence.
Generally, our map model agrees with the \textit{Spitzer} data, but there are a few differences between our model and the \textit{Spitzer} models. 
For both detectors, the JWST data display a brighter (hotter) planetary nightside.
The phase-curve offset is smaller with JWST in both cases, and the strong asymmetry present in the 4.5 \microns\ \textit{Spitzer} phase curve\change{, where the phase curve maximum occurs before eclipse but the minimum occurs after transit,} is not present in the NRS2 data \change{or our model (which could fit an asymmetry if present)}.
The dayside brightness is similar between \textit{Spitzer} 4.5 \microns\ and NRS2, but the NRS1 data show a colder dayside compared to \textit{Spitzer} 3.6 \microns. 

\begin{figure*}
    \centering
    \includegraphics[width=6.5in]{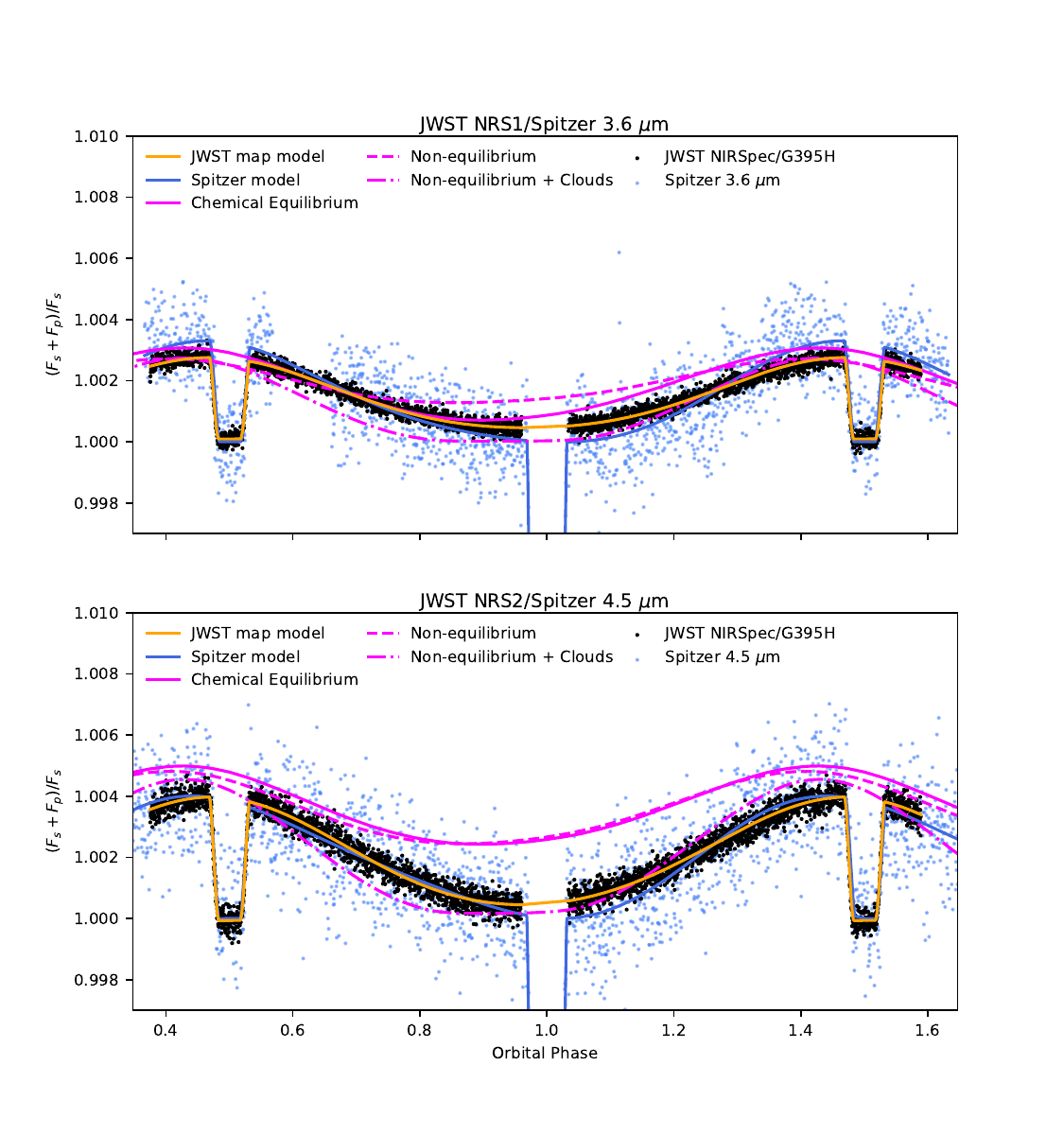}
    \caption{The NRS1 (top) and NRS2 (bottom) light curves and best-fitting map models, compared against \textit{Spitzer} phase curves from \cite{StevensonEtal2017ajWASP43bPhaseCurve}. The \textit{Spitzer} data have been binned temporally by a factor of 20 for clarity. We only show the second 3.6 \microns\ \textit{Spizter} visit, as the first is afflicted by intrapixel sensitivity variations that cannot be accurately corrected \citep{StevensonEtal2017ajWASP43bPhaseCurve}. Our data show a brighter (hotter) nightside in both bands and a dimmer (colder) dayside at the shorter wavelengths, smaller phase-curve offsets, and less asymmetry in the NRS2 phase curve than the \textit{Spitzer} 4.5 \microns\ phase curve\change{, which exhibits a phase-curve maximum before eclipse and minimum after transit}. We also show the Exo-FMS GCM light curves overplotted.}
    \label{fig:2channel}
\end{figure*}

\begin{figure}
    \includegraphics[width=3.5in]{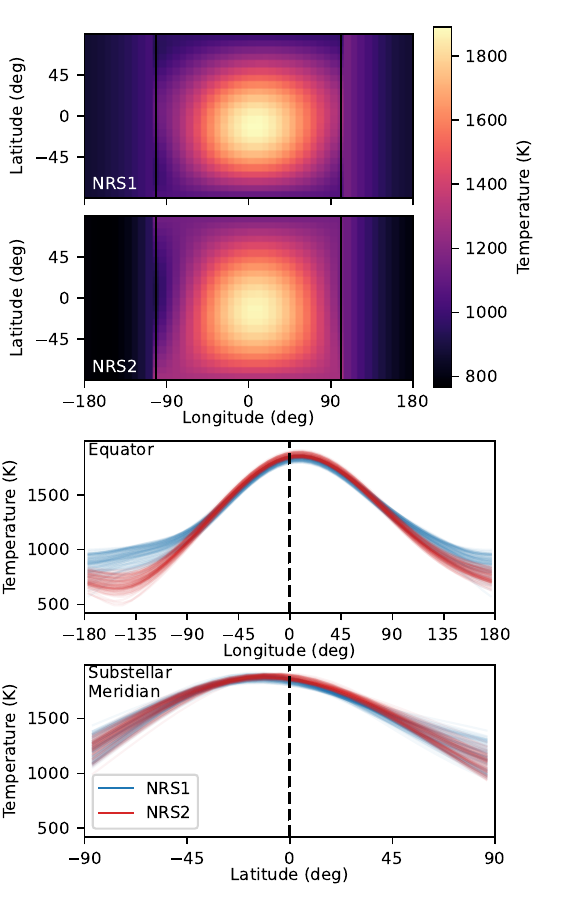}
    \caption{\changeenv{} {\bf Top:} Brightness temperature maps from fits to the NRS1 and NRS2 light curves. See the description of the inset in Figure \ref{fig:whitelight}. {\bf Bottom:} 100 samples of slices along the equator and substellar meridian from the posterior distributions of temperature maps from the fits to the NRS1 and NRS2 light curves.}
    \label{fig:2chanmaps}
\end{figure}

The maps \change{and posterior distributions of maps} associated with the light-curve fits are shown in Figure \ref{fig:2chanmaps}.
Despite the larger uncertainties, the NRS1 and NRS2 light curves require the same six eigenmaps as the white-light light curve to be fit adequately.
The best-fitting maps are broadly consistent with the white-light map (Figure \ref{fig:whitelight}), with eastward-shifted hotspots and a slight north-south asymmetry.
Relative to NRS1, the NRS2 map shows a less steep temperature gradient from the hotspot peak to the poles, but we note that the polar temperatures are highly uncertain, as the poles contribute minimally to the observed planetary flux.
The longitudes of maximum brightness are $7.7^{+0^\circ.5}_{-0^\circ.7}$ and  $8.2^{+0^\circ.6}_{-0^\circ.4}$ for NRS1 and NRS2, respectively, which are consistent with the white-light map at $\leq2\sigma$.
We calculate phase-curve offsets (see Section \ref{sec:whitelight}) of \change{$10.5^{+1^\circ.0}_{-1^\circ.1}$} and \change{$9.8^{+0^\circ.9}_{-0^\circ.8}$} for NRS1 and NRS2, respectively, which are consistent with the \textit{Spitzer} measurements, although both bands suggest slightly less efficient heat transport than previous analyses \citep{StevensonEtal2017ajWASP43bPhaseCurve, MurphyEtal2023ajWASP43bNoVariability}.
The latitudinal hotspot \change{shift} seen in the white-light analysis is still present in the NRS1 and NRS2 maps.

Depending on atmospheric composition, different wavelengths can probe different altitudes in the atmosphere and, thus, a comparison can show evidence of vertical temperature gradients.
\change{The wavelength coverage of NIRSpec/G395H is sensitive to the abundances of major atmospheric constituents H$_2$O, CH$_4$, CO, and CO$_2$, all of which could be present in WASP-43b's atmosphere based on theoretical predictions \citep[e.g.,][]{IrwinEtal2020mnrasTwoPlusD}.
If the atmosphere is not thermally inverted as expected at these temperatures \citep{HubenyEtal2003apjTiOandVO, FortneyEtal2008apjJupiters}, CO and CO$_2$ absorption could drive the NRS2 photosphere higher, reducing the temperatures in our NRS2 map. 
H$_2$O could have a similar effect in NRS1.}
Likewise, we can search for changes in the \change{longitudinal} hotspot offset with wavelength, which can tell us about the changing dynamics of the atmosphere with altitude.
In general, \change{longitudinal} hotspot offsets are predicted to decrease with increasing altitude, particularly if the atmosphere is subject to a magnetic drag \citep{BeltzEtal2022ajActiveMagDrag3D}.
Our NRS2 map exhibits a slightly larger \change{longitudinal} hotspot offset than NRS1, which could indicate that NRS2 is probing deeper in the atmosphere. 
\change{However,} our maps' \change{longitudinal} hotspot offsets are consistent at $<1\sigma$, and the temperatures of each map are similar (Figure \ref{fig:2chanmaps}), likely because the wavelength bands we are considering are broad\change{, causing the bands to probe similar pressure levels}.
Future work investigating \change{the planet's emission spectra}, spatial variations with finer wavelength sampling (e.g., in- and out-of-CO$_2$ bands to maximize vertical differences)\change{,} or fully 3D eclipse-mapping analyses \citep[e.g.,][]{MansfieldEtal2020mnrasEigenspectraMapping, ChallenerRauscher2022ajThERESA} will provide a better understanding of the atmospheric dynamics as a function of pressure.

%Further context for these temperature maps will be presented in a follow-up work that presents and analyzes the eclipse and phase-curve emission spectra.

\section{Comparison to General Circulation Models}
\label{sec:gcms}

% Elsie is slowly working on the GCM text
We compare general circulation models (GCMs) against our retrieved maps and the phase curves to study how the observation compares against theoretical predictions. 
We use the Exo-FMS GCM \citep[e.g.][]{Lee2021} with the planetary parameters taken from the MIRI LRS study \citep{BellEtal2023natasWASP43bMIRI} and assume Solar metallicity.
Due to the non-inflated nature of WASP-43b, we chose an internal temperature of T$_{\rm int}$ = 150 K.
\change{For this study,} we perform three simulations\change{:} one cloudless assuming chemical equilibrium, one cloudless with non-equilibrium kinetic chemistry (using the mini-chem module \citet{Tsai2022,Lee2023}) and \change{one with} tracer clouds following \citet{Tan2021} with non-equilibrium chemistry.
We take the best fit log-normal distribution cloud particle parameters from the MITgcm simulations from \citet{BellEtal2023natasWASP43bMIRI} (r$_{\rm med}$ =  2 $\mu$m, $\sigma$ = 2) with an amorphous MgSiO$_{3}$ composition.
The GCM T-p evolution as well as chemical and cloud tracers are coupled to a mixing length theory \citep[e.g.][]{Ludwig1999} convective scheme, with tracer convection emulated through a vertical diffusive component to the transport the same as in \citet{Lee2024}.
All models use correlated-k radiative-transfer and include radiative-feedback from the changing volume mixing ratio (VMR) of species as well as cloud opacity feedback.
We use a stellar spectrum from the PHOENIX \citep{Husser2013} stellar model grid ($T_* = 4520$,  [M/H] = 0.0, and log($g$) = 4.63).
Overall, this model is similar in set-up to the directly imaged exoplanet simulation in \citet{Lee2024}, but modified for hot Jupiters to include shortwave irradiation from the host star.

To post-process the GCM simulation output, we use the gCMCRT code \citep{Lee2022} and produce white light phase curves for the NIRSpec G395H NRS1 and NRS2 wavelength ranges for a direct comparison to the observational data (Figure \ref{fig:2channel}). 
We take the same chemical and cloud properties from the GCM output as input to gCMCRT.

% Ryan's attempt at some text -- Elsie feel free to modify
Despite not being specifically tuned to match the data, the GCMs follow the trend and magnitudes of the phase curve quite well, especially when clouds are included.
In NRS1, the measured dayside emission is well-matched by all three models, while the nightside flux is slightly overestimated by the chemical-equilibrium model and slightly underestimated by the non-equilibrium model with clouds. 
In NRS2, all three models slightly overestimate the dayside flux, but the cloudy model provides a close match for the nightside emission, in agreement with \cite{BellEtal2023natasWASP43bMIRI}.
The GCMs have \change{phase-curve} offsets of 25 -- 32$^\circ$, all of which overestimate our measured \change{phase-curve} offset of $\sim\,10^\circ$. 
This could indicate the presence of atmospheric drag restricting heat transport that is not included in the GCMs, such as interactions with a magnetic field \citep{BeltzEtal2022ajActiveMagDrag3D}.
Higher metallicities can also reduce the \change{phase-curve} offset through changes in the heat capacity and specific gas constant \citep{KatariaEtal2015apjW43bGCMs, ParmentierCrossfield2018haexPhaseCurves}.

\section{Conclusions}
\label{sec:conclusions}

We presented an analysis of a JWST NIRSpec/G395H phase curve of the canonical hot Jupiter WASP-43b.
\change{
With the full phase curve, we use the planetary transit to constrain WASP-43b's orbital parameters, we use the phase-curve variation obtain tight constraints on the planet's longitudinal brightness distribution, and we use the eclipse ingress/egress to determine the detailed brightness distribution of the planet's dayside, including the first significant evidence ($4.0\sigma$) of a latitudinal hotspot offset in an exoplanet atmosphere.
}

By modeling the white-light transit, we improved constraints on the planet's orbital inclination, semimajor axis, and transit time by a factor of $\sim\,2$, and provide a precise measurement of the planet's radius.
These improvements in the system parameters led to significantly increased confidence in our mapping analysis.
Future observing programs with an eclipse-mapping element should consider whether system parameters are known well enough to reach their mapping goals, and, if not, consider including a transit observation.

We fit a 2D planetary brightness model to the phase-curve variation and both eclipses, simultaneously. 
Our white-light map shows a peak dayside brightness temperature of nearly 2000 K, dropping to an average of $\sim1000$ K on the nightside.
We measured a longitude of maximum brightness and phase-curve offset consistent with previous measurements with \textit{Spitzer} and JWST MIRI/LRS.
Maps fit to the NRS1 and NRS2 light curves are similar, with consistent hotspot \change{locations}.
However, we note that we find hotter nightsides than the \textit{Spitzer} observations in both bandpasses, and we do not see the asymmetry present in the \textit{Spitzer} 4.5 \microns\ phase curve \citep{StevensonEtal2017ajWASP43bPhaseCurve}.

In both the white-light and single-detector maps, we see evidence for a latitudinal hotspot offset \change{that is only confidently detected with our transit-refined orbital parameters}.
Such an offset could be created by a dipole magnetic field misaligned with the planet's rotational axis interacting with an ionized atmosphere, a somewhat unlikely scenario given WASP-43b's temperature regime.
Patchy clouds, which could form on the dayside of WASP-43b, could also create spatial asymmetries, although observational evidence suggests a cloud-free dayside atmosphere \citep{BellEtal2023natasWASP43bMIRI}.

Through comparison with GCMs, we see clear evidence for clouds on the nightside of the planet, particularly in NRS2, where the clear models are easily ruled out as too bright.
Also, the measured longitudinal hotspot offsets are lower than those predicted by the GCMs, potentially indicating unmodeled atmospheric drag or a higher-than-expected metallicity.
Magnetic effects could explain both the nonzero latitude and closer-to-substellar longitude of the hotspot.

This dataset will provide a wealth of information in future work.
The transmission spectrum will provide insight into the extent of clouds.
Study of the eclipse spectra will give detailed information on the vertical thermal structure and chemical inventory of the planet's atmosphere, giving additional context to the maps presented here.
Atmospheric retrievals on the phase-resolved emission spectra and 3D eclipse mapping will give constraints on the multidimensional atmospheric properties. 

\begin{acknowledgments}
\change{We thank the anonymous reviewer for their insightful comments that led to the improvement of this manuscript.}
We thank contributors to SciPy, Matplotlib, Numpy, and the Python Programming Language.
E.K.H. Lee is supported by the SNSF Ambizione Fellowship grant (\#193448).
N. OG is supported by the Davidsen Fellowship awarded by the Space Telescope Science Institute.
\end{acknowledgments}

\software{NumPy \citep{HarrisEtal2020natNumPy}, Matplotlib \citep{Hunter2007cseMatplotlib}, SciPy \citep{VirtanenEtal2020natmSciPy}, Scikit-learn \citep{PedregosaEtal2011jmlrScikitLearn}, starry \citep{LugerEtal2019ajStarry}, ThERESA \citep{ChallenerRauscher2022ajThERESA}, MC3 \citep{CubillosEtal2017ajRedNoise}, BATMAN \citep{Kreidberg2015paspBATMAN}, FIREFLy \citep{RustamkulovEtal2022apjlFIREFLy, RustamkulovEtal2023natWASP39b}, gCMCRT \citep{Lee2022}.}

\facility{JWST (NIRSpec)}

\appendix

\section{Correlated Noise}
\label{sec:corrnoise}

\begin{figure}
    \centering
    \includegraphics[width=3.25in]{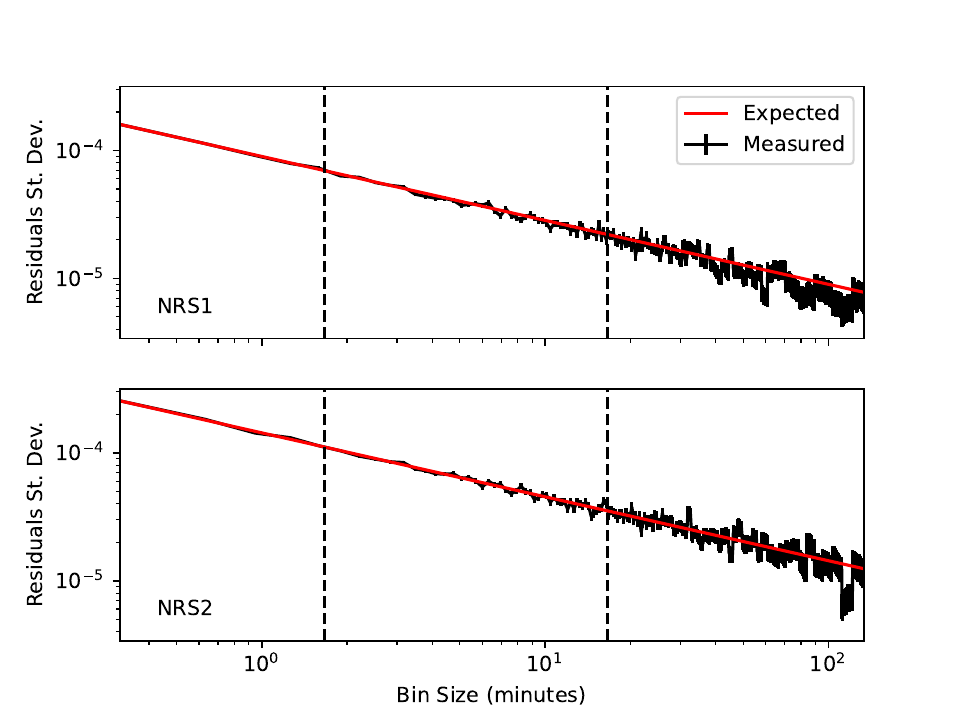}
    \caption{Allan variance plots of the NRS1 (top) and NRS2 (bottom) light-curve residuals. The black line shows the standard deviation of the model residuals with uncertainties and the red line shows the expected $1/\sqrt{N}$ trend. If correlated noise is present at a given timescale, the black line would be above the red line. For both detectors, the black line follows the red line, showing no significant residual correlated noise. The vertical dashed lines mark a region between the ingress/egress time and a tenth of the ingress/egress time, approximating the range of timescales where correlated noise could interfere with our mapping signal.}
    \label{fig:allan}
\end{figure}

We note that eclipse-mapping analyses can be particularly susceptible to biases caused by high-frequency correlated noise. 
Unlike traditional eclipse spectroscopy, which is concerned with signals the duration of the entire eclipse, eclipse mapping relies on signals that occur during ingress and egress. 
For WASP-43b, with its extremely short orbital period, the ingress and egress duration is $\approx17$ minutes. 
We computed the Allan variance of the residuals for both detectors and find no evidence for correlated noise at any timescale (Figure \ref{fig:allan}).

\bibliography{wasp43b.bib}

\end{document}